\documentclass[12pt,preprint]{aastex}

\shorttitle{The Variable Reflection Nebula Cepheus A East}
\shortauthors{Hodapp,Bressert}

\begin{document}

\title{THE VARIABLE REFLECTION NEBULA CEPHEUS A EAST}
\author{Klaus W. Hodapp\altaffilmark{1}, Eli Bressert\altaffilmark{2}} 

\altaffiltext{1}{
Institute for Astronomy, University of Hawaii,\\
640 N. Aohoku Place, Hilo, HI 96720,
\\email: hodapp@ifa.hawaii.edu }

\altaffiltext{2}{
Harvard-Smithsonian Center for Astrophysics,\\
60 Garden Street, MS 67, Cambridge, MA 02138,\\
email: ebressert@cfa.harvard.edu}

\begin{abstract} 

We report K$^\prime$-band imaging observations 
of the reflection nebula associated with Cepheus A East
covering the time interval from 1990 to 2004.
Over this time the reflection nebula shows
variations of flux distribution, which we
interpret as the effect of inhomogeneous and varying extinction in
the light path from the illuminating source HW2 to the reflection
nebula. The obscuring material is located within typical distances
of $\approx$ 10~AU from the illuminating source.

\end{abstract}


\section{INTRODUCTION}

\subsection{Variability in Young Stars}

Variability was one of the characteristics originally used to define the first
optically discovered class of young stars, the T Tauri stars \citep{joy1945}.
It is therefore no surprise that infrared observations of young
stars also commonly show variability. An early study by \citet{cohen1976} showed
that typical young stars are variable at all wavelengths from UV to the
thermal infrared, with often pronounced color effects.
A comprehensive study of variable stars in the Orion~A molecular cloud
by \citet{carpenter2001} demonstrated that about half of all stars in that
region are variable, and that most of the variable stars are indeed young.
Most young variable stars vary on timescales of days, consistent with
stellar rotation modulation of the star's light, while only a small minority
of these stars shows variability suggestive of accretion instabilities,
even though the latter mechanism provides some of the most spectacular examples of
variable young stars (e.g., \citet{herbig1977} and \citet{hod99}).

\subsection{Variable Reflection Nebulae}

Stars form out of molecular cloud material and young
stars in the early phases of their evolution, in particular Class I and
some Class II objects, are therefore usually associated with reflection nebulosity. In most
cases, the reflection nebulosity is formed by the illuminated walls of the cavity created by the
outflow from the young star, and is therefore outside of the plane of the protoplanetary
disk surrounding many young stars. Generally, this reflection
nebulosity can therefore be expected to also show variability 
in response to changes in the overall brightness
of the young star. However, given the location of the reflection nebula
perpendicular to the disk plane, and therefore also in the polar direction of 
stellar rotation, variability caused by rotational modulation by star spots
is not expected to be seen in the typical "cometary" or bipolar reflection nebulae
associated with low-mass young stars.

Variability of the reflection nebulosity around some stars has been noted
as far back as the mid 19th century using visual observations. The earliest
example appears to be Hind's variable nebula near T Tauri that was discovered
in 1852 and was easily visible even in relatively small telescopes in 1855. 
However, as summarized in detail by \citet{barnard1895}, it appeared much fainter in 1861
when only traces of it were visible even in
the largest telescopes available at that time.
The variability of the nebula around R CrA that was first noted in 1890 and was discussed
in detail by \citet{knoxshaw1916}, who noted morphological changes on minimum timescales
of one week, uncorrelated with changes of the brightness of the illuminating star R CrA.
The morphological variability of Hubble's Variable Nebula (NGC 2261) was reported
by \citet{hubble1916} and documented with photographic records. 
Its variability was successfully explained by \citet{lampland1926}
as being caused by variations of the illumination of an essentially constant screen
of scattering material in the reflection nebula.

The study of the variability of even younger, more deeply embedded objects is still
in its infancy, because infrared observations are just now establishing a sufficiently
long historical record to allow such studies.
An infrared study of the L~483 reflection nebula using data from the same K$^\prime$ survey follow-up
and very similar data reduction techniques is being published by \citet{connelley2008}.
Despite its lower mass, L~483 shows overall similar photometric variations to the ones reported here for Cepheus~A East.

\subsection{Cepheus~A~East}
Cepheus~A~East (Cep~A~East) at a distance of 725~pc \citep{blaauw1959} was historically the second object
discovered to show a large bipolar molecular
outflow of complex structure \citep{rodriguez1980}.
The main energy source of the entire Cep~A region is the source HW2 found by
\citet{hughes1984} by continuum radio interferometer observations. High spatial resolution VLA measurements
by \citet{curiel2006} demonstrated the proper motion of up to 480 km s$^{-1}$ of the radio (3.6 cm) jet emanating from
HW2 and tentatively identified the driving source of the jet with a 7~mm continuum emission
knot.
\citet{curiel2006} pointed out that, 
while morphologically somewhat similar to the familiar jets of low-mass YSOs,
the HW2 jet is faster, has higher radio flux
and a different spectral index from low-mass jets. Water masers associated with
this source are comparatively luminous and the driving source luminosity is high,
all pointing to the fact that HW2 is a young high-mass star and the source
primarily responsible for both the high-velocity jet and the large-scale outflow
in Cep~A \citep{narayanan1996}.

Deep, continuum-subtracted images in the H$_2$ 1--0 S(1) line by \citet{cunningham2006}
showed the bipolar outflow of Cep~A~HW2 in great detail and allowed the identification of several
shock fronts that probably resulted from separate episodes of jet activity from a
precessing outflow source. In addition, \citet{cunningham2006} found a second, undisturbed,
straight outflow. 
This second outflow may originate from the very deeply embedded radio source HW3c 
and the associated sub-millimeter sources 
HW3c-SMA and SMA4 that are located 4$\arcsec$ to
the south of HW2. 
The positions of these outflow sources are indicated in Fig.~1.
Another possible source of this outflow will be discussed in Section 4.

Polarimetric measurements in the L-band by \citet{lenzen1984}
demonstrated the high degree of polarization of the Cep~A~East reflection nebula and
showed that the illumination originates very close to HW2, a result confirmed by J, H,
and K imaging polarimetry by \citet{casement1996}. The high resolution polarization map of
\citet{jones2004} also identified HW2 as the dominant source of illumination, even though
illumination by other, fainter sources is also evident.

The highest spatial resolution radio observations of HW2 to date were obtained by \citet{jimenez2007}
and showed, at sub-arcsec resolution in SO$_2$ line emission, a 600~$\times$~100~AU extended structure that
they interpreted as evidence for a rotating disk of radius
300~AU and a mass of 1~M$_\odot$ around the source of the radio jet, HW2. In addition, 
their 7~mm radio continuum map showed emission not only in the direction of the well-known HW2 jet,
but also along the walls of the outflow cavity, the same interface region between the disk and the
jet where
\citet{torrelles1996} 
found water maser emission. 
\citet{jimenez2007} interpreted this as evidence for ongoing
photo-evaporation of the collimating disk of the radio jet.

However, sub-millimeter observations with the Submillimeter Array (SMA) by \citet{brogan2007} cast doubts on
whether the disk around HW2 does indeed extend out to a radius of 300 AU.
They labeled the north-western part of the extended SO$_2$ emission seen by \citet{jimenez2007}
as a separate source (SMA2) and pointed out 
the differences in thermal and chemical
properties between the HW2 and SMA2 sources.
In this scenario the region around
HW2 is a small cluster of young stars,
sufficiently dense for the induced merger
hypothesis for the formation of massive stars proposed by \citet{bonnell2005} to work.
At this time, while the existence of a collimating disk of the HW2 outflow is strongly
supported and while this disk is likely undergoing photo-evaporation, 
the question remains open whether the more extended (about 1 arcsec) sub-millimeter flux near HW2 originates from a single
large (diameter 600~AU) disk around Cep A HW2 or from a number of individual sources in a dense cluster.

\section{OBSERVATIONS AND DATA REDUCTION}
\subsection{Data Acquisition}
The observations reported here were collected over the span of 14 years using
two different infrared cameras at the UH 2.2m telescope. The observing dates are summarized
in Table~1.
The earliest images of the Cep~A reflection nebula used in this study were taken
as part of the K$^\prime$ imaging survey of outflow sources 
\citep{hod94} on September 8, 1990 (UT). The instrument
used for these observations was the 256$\times$256 UH-NICMOS-3 camera \citep{hodapp1992} 
in its wide field mode with
0.75$\arcsec$ pixel$^{-1}$ image scale at the f/10 Cassegrain focus of the UH 2.2m telescope.
The other data were taken between 1998 and 2004 as part of a program to monitor some of
the regions in the original K$^\prime$ survey for variability. For these follow-up observations,
the UH QUIRC camera \citep{hod96} was used, a 1024$\times$1024 camera with a HAWAII-1 HgCdTe detector
array fabricated by Rockwell (now Teledyne Imaging Systems).

All the observations reported here were done in the K$^\prime$ filter \citep{wainscoat1992} 
at an effective wavelength of
2.15$\mu$m.
Both detector
arrays were based on the Rockwell 2.5$\mu$m PACE-1 HgCdTe material and the cameras used
very similar optical designs, so that the effective wavelengths
of both cameras were closely identical. The data acquisition procedure was also very
similar for both cameras. The position on the object itself was covered by a total
of 6 dithered exposures for the UH-NICMOS-3 observations, and 10 for the QUIRC observations.
Sky frames were taken to the east and west of the central position. 
We took 3 images in either sky position in 1990 with the UH-NICMOS-3 camera 
and 5 sky images in the later QUIRC observations,
with some overlap with the central position to allow the relative alignment of the individual
frames.
Differential domeflats were used to generate the flatfields and bad pixel masks.
Skyframes were computed by median filtering the stack of ''sky'' position frames
east and west of the object. The flatfielded and sky-subtracted images were then
registered and co-added. 
The QUIRC images taken between 1998 and 2004 typically have a 5 $\sigma$ limiting
magnitude of K$^\prime$~=~16.5.

\subsection{Data Reduction}
Variations of extended objects are best studied using difference images.
In preparation for computing the difference images, the UH-NICMOS-3 image from 1990 was
magnified to roughly match the pixel scale of the later, more finely resolved QUIRC images. 
The images from
the different epochs were then roughly aligned by shifting. Fine position-, scale-, 
rotation-, and PSF-matching were done with the IRAF tasks geomap, geotran, and psfmatch.

The sky background in the K$^\prime$ filter has a large component of time-variable OH emission.
The original frame alignment process matched the sky level of adjacent frames, but
leaves the absolute sky level of the resulting mosaic image poorly defined. 
Therefore, this sky level was measured in several
empty areas of the images close to the main Cep A East reflection nebula, and then
subtracted to produce an image without artificial signal pedestal. 
While all the images were taken under nominally
photometric conditions with the same integration time and data acquisition process,
we nevertheless checked for photometric consistency, because, on timescales of years, 
the reflectivity
of the telescope mirrors will change periodically with the re-aluminization cycles. 
A set of stars without
noticeable variability was selected and the difference in the average instrumental
magnitudes of this set of stars was computed for each image. The individual images
were then corrected for the small (less than 3\%) photometric calibration difference between
the images. 

One set of QUIRC images was PSF-matched to the low-resolution 1990 data, and
the average of these 1998 to 2004 QUIRC images was used to produce a low-resolution
time-averaged image representative of the average flux in Cep~A. This low-resolution
average image was then subtracted from the 1990 UH-NICMOS-3 camera frame.
To preserve the higher spatial resolution of the QUIRC images, the same
time average from 1998 to 2004 was computed for a version of the QUIRC images that were only slightly
smoothed to even out the small PSF differences among those data. This high-resolution
version of the time-averaged image was then subtracted from all the QUIRC images (from 1998 - 2004).

The best of the QUIRC images, with a FWHM of 0.65$\arcsec$ is shown in Fig.~1
and illustrates the knotty and filamentary morphology of the reflection nebula,
showing detail down to the resolution limit of this image.
All astrometry in this and the other figures is based on the 2MASS catalog \citep{skr06}.

In Fig.~2, we summarize the results on the morphological changes in the 
Cep~A~East reflection nebula. In the left column of this figure, we show the
direct K$^\prime$ images in the common photometric calibration described above. 
The difference images computed are shown in the center column of Fig.~2 and best illustrate
the changes in the light distribution in the Cep~A~East reflection nebula.
Since the variations in the reflection nebula appear to be caused by
shadows cast into the material in the outflow cavity of Cep~A, we divided
the difference image by the original flux distribution, to
distinguish between the spatial structure of the reflection nebula and the
shape of the shadows. These normalized difference images are shown in the right
column of Fig.~2.

To better illustrate the amplitude of the changes, we chose a set of rectangular regions
covering distinct features of the reflection nebula, shown superposed on one
of the images in Figs.~1~and~3. Using the IRAF task ''imstat'', we computed the signal
average in these rectangular areas after subtracting a similarly measured
small sky value. The resulting light curves are plotted in logarithmic
flux units (decadal logarithm of the averaged signal in digital units), in Fig.~3.
Note that the two easternmost integration boxes are located outside of the field
covered in Fig.~1.

\section{DISCUSSION}

\subsection{Shadow Effects in Cepheus~A East}
The changes
in the brightness of different features in the nebula are noticeable in
the left (direct image) column of Fig.~2, but are not
prominent. The direct images mainly serve to re-emphasize the point made by
earlier authors for similar objects, e.g., \citet{knoxshaw1916} that the observed variations are
not related to movements in the directly observable part of the nebula, or to physical 
changes in the scattering material.
The variations in the Cep~A~East reflection nebula are prominently visible in
the difference images shown in the center column of Fig.~2.
Note that in all of Fig.~2, dark tone denotes higher signal. Shadows therefore
appear white in this figure.
The right column of Fig.~2 shows the difference images divided by the
original image, to better separate the illumination and shadowing effects from
the density inhomogeneities of the reflection nebula. A uniform illumination
change of an inhomogeneous reflection nebula would show a uniform signal in
the right column.
This sequence of images clearly demonstrates that
the changes in different parts of the reflection nebula are correlated. 

One arcmin of projected distance in Cep~A (725 pc distance) corresponds to
a distance of 0.69 light-years. The maximum projected extent of the Cep A
reflection nebula ($\approx$3$\arcmin$) is about two light-years, 
and the typical size of areas
with distinct light curves is as small as a few arcseconds or about a light month.
In the well sampled time interval from 1998 to 2004, changes occur on a 
timescale of several years. 
In many areas of the reflection nebula, the brightness returns to roughly
the initial (1998) brightness after the time interval of 6 years.
Changes in the illumination pattern
of the reflection nebula are therefore not dominated by light travel time
effects, even though light travel effects are not completely negligible. 
The individual image that appears most consistent with an effect of 
light travel time is the 1999 data point, where the parts of the nebula
farther away from the illuminating source are fainter than those closer
to it. 

The other images show the pattern expected from shadows being projected
into the reflection nebula: elongated patterns pointing towards the 
illuminating source. This is most prominent in the 1990 and 2004 images,
but is also the dominant effect in the 1998, 2000, and 2002 images.
The images from 1998 and 2002 show brightness changes at the southern 
and western edge of the reflection nebula that are different from those
in the bright, main parts of the reflection nebula. This can best be
explained by difference in the shadowing along the line of sight to
the observer.

We show the light curves of six selected areas of the reflection nebula
in Fig.~3. These areas were simply selected to represent regions of the reflection
nebula at different distances from the illuminating source where rectangular
integration boxes were not significantly contaminated by stars.
We note that over the 14 year span of our observations, the overall brightness
of the reflection nebula has not changed significantly. The individual
selected areas of the reflection nebula have distinct light curves, even
though some common features are noticeable. All areas show an increase in
brightness, albeit at different magnitudes, between 1990 and 1998, and a
decrease in brightness between 1998 and 1999. Aside from these similarities,
the light curves differ in the details. For half of the areas (A, E, and F)
the decline in brightness continues for the December 2000 data point, while for the
others, the brightness increases between 1999 and 2000. Between 2002 and 2004,
two areas show an increase in brightness (A and E), three show a decline (B, D, and F),
while area C is nearly constant.
The differences can be interpreted on the basis of shadows moving across
the reflection nebula. Restricting the discussion to the well-sampled time
interval from 1998 to 2004, we note that in 1998, the north-west half of the
reflection nebula is brighter than the south-eastern half, while the situation
is reversed in 2000. Between 2000 and 2004, the illumination pattern reverses
again, leading to a situation where the north-western half of the nebula is
brighter and the south-eastern half fainter. The light curves are indicating
a phase shift of about one year between areas A and E on the north-western
side of the reflection nebula's symmetry axis, and B, C, D, and F on the south-eastern
side, with the north-western areas reaching the minimum one year after the
south-eastern points. 
We interpret this as evidence that a shadow sweeps from the south-east to the
north-west across the reflection nebula.

Our data appear to adequately sample the illumination changes with time.
We do not see erratic data points in the light curves in Fig.~3 that
would indicated under sampling of the light curve.
Our data do not allow the determination of a period for the brightness variations, if
such a periodicity does indeed exist. All selected areas suggest, however, that
we may have observed at least one minimum between 1998 and 2004.
For the purpose of estimating the location of the obscuring material, 
we summarize the results as a full reversal of the light curves over
a 6 year time interval and across a 120$\degr$ opening angle of the reflection
nebula.

The simplest model to explain the basic results is that of shadows
being projected onto a screen, the reflection nebula, that has a complex structure.
In the difference images against the time-average, the dividing line separating
positive and negative regions is often nearly straight, and pointing to the
illuminating source at or very near the position of HW2.
This suggests that the
changes in illumination are caused by obscuring clumps of material close
to the illuminating source HW2 and that the changes in the illumination of the
nebula are shadows cast by these clumps. 
Expanding on that simple model, we clearly see additional effects, most notably
in the 1999 and 2002 data, where the areas close to the illuminating source
show different photometric behavior than those further away. In 1999, the 
southern part of the near parts of the reflection nebula have lower flux
than the rest of the nebula, while in 2002, the northern part of that nebula
shows similar behavior. These differences between the near and the far areas of
the reflection nebula could be explained by light travel effects, but, of course,
may also be caused by the three-dimensional structure of the reflection nebula.

\subsection{Location and Properties of the Obscuring Material}

As part of their study of the morphological evolution of bipolar outflows based
on a comparison of Spitzer data and SED models, \citet{seale2008}
have shown that the cavities of bipolar nebulae not only widen over time, but also
decrease in density. For a very young object with strong outflow activity like
Cep~A~HW2, the density within the cavity must be high. For YSOs with ages of 10$^5$
years, \citet{seale2008} find typical densities in the cavity of 10$^{-20}$ gcm$^{-3}$.

The hydrodynamical models by \citet{cunningham2005} of cavity excavation in the
outflows of massive stars indicate that the boundary between the fast moving
wind or jet and the cavity walls is expected to be turbulent. 
Their models also show that the leading shock front rapidly fragments and becomes
clumpy, which is relevant to the discussion of Cep~A since there is evidence
\citep{curiel2006}, discussed in more detail below, that 
the driving source (HW2) of Cep~A is not constant, but repetitive
on timescales of a few years.
On larger spatial scales and longer time scales, \citet{bally2008} finds
evidence that the HW2 outflow source precesses and ejects material into
different directions approximately every 2200 years.
Therefore, due to this wandering of the jet and its repetitiveness, each new
''leading'' shock front exhibits the same tendency for rapid fragmentation.
Matching this model, the appearance of the Cep~A~East reflection nebula on near-infrared images is
very filamentary and knotty, with the smallest features being unresolved in
our best images with a FWHM of 0.65$\arcsec$ shown in Fig.~1.

We note that the changes in the
illumination pattern, i.e., the peak-to-peak variations of the light curves in Fig.~4, are of
the order of a factor of 1.6 for the brighter regions of the reflection nebula, and up to about
a factor of 2 for the more distant regions. 
This can be explained if
these obscuring clouds are far from opaque at the observed near-infrared wavelengths, 
or if they only affect a small fraction
of the depth of the nebula. 
The variations in the reflection nebula's surface brightness correspond to 0.5 to 0.75 magnitudes of K$^\prime$ band
extinction, if the obscuring clouds shadows
the full depth of the reflection nebula. 
Extinction in the optical would be approximately an order of magnitude larger, even
though the interstellar extinction law (e.g., \citet{rieke1985}) is probably not
applicable so close to a protostar.
This is consistent with the fact that some
of the variable reflection nebulae historically observed at optical wavelengths had
shown stronger, more easily recognized variations.

In Fig.~5, we show the normalized differential image based on the
2004 data with contours of the normalized brightness difference superposed.
The data are the same as those in the lower right panel of Fig.~2, but for
Fig.~5 we only show the area of the clearly visible shadow effect.
The superposed contour show that the shadow is a gradual transition rather
than a sharp edge.
The position of the illuminating source HW2 is also indicated.
The main result illustrated by Fig.~5 is that the profile of the light variations
is smooth, indicating a rather smooth distribution of the obscuring matter.

For an estimate where the obscuring cloud condensations are located, we consider the
various velocities involved in the outflow phenomena in Cep~A. The radio jet
shows several condensations that have projected proper motions, measured with the VLA by
\citet{curiel2006}, of about 480 km s$^{-1}$, or spatial velocities in the range between
525 and 650 km s$^{-1}$. These speeds are higher than those of low-mass YSOs, but are
in general agreement with those found in other high-mass young stars. By tracing 
several condensations in the jet, \citet{curiel2006} also find evidence that the
driving source of the jet undergoes major mass ejections every 1.85 yr and is 
probably precessing and/or nutating, judging from a slight asymmetry and wiggling
of the radio jet.
For the purpose of explaining the shadowing effects, the jet velocity along the
jet axis, coinciding roughly with the symmetry axis of the outflow cavity, is
not relevant. Varying shadow effects can only be produced by motions perpendicular
to this axis.

Masers, in particular H$_2$O masers, can be used to trace the interface between
the molecular outflow and the ambient cloud material. The H$_2$O masers in Cep~A
have recently been used by \citet{gallimore2003} to map out an expanding ring of maser emission near the
HW2 driving source of the outflow. They found that the expansion velocity has,
over the time interval from 1996 to 2000, decreased from 30 - 40 km s$^{-1}$ to
$\approx$ 13 km s$^{-1}$. \citet{vlemmings2006} have further concluded that 
by 2004, this expansion has effectively stalled. 
Molecular emission lines of various species measured by \citet{brogan2007} 
have widths of the order of 10 to 15 km s$^{-1}$, indicating typical motions in
the line of sight, outside of the outflow itself. 

As discussed by \citet{sonnentrucker2006}, the discovery of gas-phase CO$_2$ emission
in Cep~A East constrains the shock speeds in that region to between 15 km s$^{-1}$, the
minimum necessary for efficient sputtering from solid CO$_2$, and 30 km s$^{-1}$, 
where CO$_2$ would be destroyed by reaction with hydrogen.

As an additional line of arguments, we can also assume, as a simplifying assumption, 
that the obscuring objects, most likely dusty clumps,
are in Keplerian orbits around the illuminating star.
For a rough estimate of the speed with which these obscuring clouds transit in
front of the illuminating source (HW2), we observe that from 1998 to 2004,
our light curves suggest that a full transit of some obscuring feature has
occurred. In 2004, the light curves in most positions have returned to roughly
the 1998 levels, and the overall light distributions are similar, in that the north-west
half of the reflection nebula is bright, and the south-eastern half is darker. 
As an order of magnitude estimate, we assume that over the 6 years of well-sampled
data, an individual obscuring clump has traveled the equivalent of the reflection
nebula opening angle relative to the illuminating star, roughly 120$\degr$. 
Assuming as an extremely simplified model that the obscuring clouds are in
Keplerian motion around HW2, and assuming its mass to be 9 M$_\odot$
\citep{jimenez2007},
this orbital period of $\approx$18 yrs corresponds to
an orbital radius of 14.3 AU, comparable to the orbit of Uranus in our own solar
system, and to a circular orbital velocity of 24 km s$^{-1}$. 

The various measurements of velocities outside of the radio jet agree within a
factor of two with the Keplerian orbital velocity corresponding to our above
estimate for the crossing time of the obscuring clouds over the angular extent
of the reflection nebula. 
While it is unlikely that the Keplerian motion model
describes the motion of dust clouds in the turbulent outflow cavity adequately, 
the general agreement of this orbital velocity estimate with the observed
velocities of gas components in the outflow cavity is encouraging.
If we assume again that the obscuring clouds travel along a 120$\degr$ arc of
a circle across the illuminating path of the reflection nebular over the time
span of 6 years, the radius of this arc must be 6 AU for a velocity of 10 km s$^{-1}$
and 18 AU for 30 km s$^{-1}$. This shows that the obscuring clouds are located 
at typical distances from the illuminating source HW2 corresponding to the orbital
radii of the outer planets of our Solar System.

The obscuring clouds responsible for changing the illumination of the
reflection nebula can obviously not be located in the plane of the disk collimating
the outflow, but must be located in the area of the outflow
cavity. Our results show that in the case of Cep~A~HW2, the outflow cavity is not
empty, but that within distances of order of tens of AU, significant quantities of absorbing
material cross the light path in the direction of the reflection nebula.

\section{THE BIPOLAR NEBULA HW7}
The small bipolar nebula shown (highlighted) in Fig.~1 lies on the axis of the
second, straight outflow in Cep~A. It was first noted by \citet{goetz1998} in
narrow-band images, but interpreted as a partly obscured bow shock. \citet{bally2008},
citing \citet{cunningham2006} shows that this nebula lies on the symmetry axis
of the outflow that terminates in HH 168 to the west, and a faint H$_2$ bow shock
in the east. This outflow is usually thought to originate in the radio source HW3c
(indicated in Fig.~1),
which shows all the signposts of an energetic outflow and also lies on the
symmetry axis of the flow. Unfortunately, HW3c is too deeply embedded to be
visible at any wavelength shortward of radio wavelengths, and its association
with HH 168 is therefore hard to ascertain. 

In Fig.~6, we compare publicly available Spitzer IRAC images at the position of HW7 obtained
originally by G. Fazio with the K$^\prime$ image combined from all our QUIRC data. 
Within the limited angular resolution of the Spitzer data, the short wavelength channel
shows features similar to the K$^\prime$ image, while at longer wavelengths, the object
looks diffuse without a pointlike central source. Since the K$^\prime$ image suggests that
the small bipolar nebula is seen nearly edge-on, it is indeed to be expected that
the embedded source in the nebula does not become directly visible even at 8$\mu$m.
In Figs.~1 and 6, we have marked the positions of the radio-continuum emission maxima
HW7c and HW7d on our image. These radio sources coincide closely with the two lobes of
the near-infrared bipolar nebula. The source GPFW2 that \citet{goetz1998}
had found in the L and M bands is also seen on our image in Fig.~1 and the position of HW7a,
which is close, but not coinciding with that point source, is marked in Fig.~6.
On the radio maps of \citet{hughes1984}, a small
unnamed knot of radio emission coincides with the central position of the bipolar
nebula.

While our image and other near-infrared data, e.g. \citet{goetz1998}, certainly
suggest that this bipolar nebula could plausibly be identified as the driving
source of the HH 168 outflow, those arguments are not in themselves conclusive either. 
Polarization measurements could, in principle, be used to determine if the 
small bipolar nebula is indeed internally illuminated. Unfortunately, the
polarization maps of \citet{casement1996} do not have the spatial resolution
to discern a centro-symmetric pattern in that small nebula.
The uncertainty over the location of the second most important outflow source
in the Cep~A complex clearly requires future studies.

\section{CONCLUSIONS}

We have presented K$^\prime$-band imaging data of the Cep~A East reflection
nebula spanning the time from 1990 to 2004. We find that the reflection nebula
shows surface brightness variations on timescales of years in a manner that 
is consistent with shadowing from
obscuring clouds of material close to the illuminating source.
The timescales of the variations in the reflection nebula, combined with typical
velocities measured in gas in the vicinity of HW2 allows the conclusion that
the obscuring clouds are located close to the illuminating source, at distances
comparable to the outer planet orbits in our own Solar System. The shadows seen projected
into the reflection nebula are not very sharp, suggesting that the obscuring
clouds have smooth profiles without a well-defined edge and only moderate optical
depth at the observed near-infrared wavelength.

\acknowledgments

This work is based in part on observations made with the
Spitzer Space Telescope, which is operated by the Jet
Propulsion Laboratory, California Institute of Technology
under a contract with NASA.

This publication makes use of data products from the
Two Micron All Sky Survey, which is a joint project
of the University of Massachusetts and the
Infrared Processing and Analysis Center / California
Institute of Technology, funded by the
National Aeronautics and Space Administration
and the National Science Foundation.

We wish to thank the referee, John Stauffer, for his constructive comments.

\clearpage
\begin{deluxetable}{ccc}
\tabletypesize{\scriptsize}
\tablecaption{Log of Observations}
\tablewidth{0pt}
\tablehead{
\colhead{Date} & \colhead{Instrument}
}
\startdata
8 Sept. 1990 & UH NICMOS3\\
29 Sept. 1998 & UH QUIRC\\
19 Sept. 1999 & UH QUIRC\\
9 Dec. 2000 & UH QUIRC\\
1 Aug. 2002 & UH QUIRC\\
31 Jul. 2004 & UH QUIRC\\
\enddata
\end{deluxetable}

\clearpage
\begin{figure}
\figurenum{1}
\includegraphics[scale=0.8,angle=0]{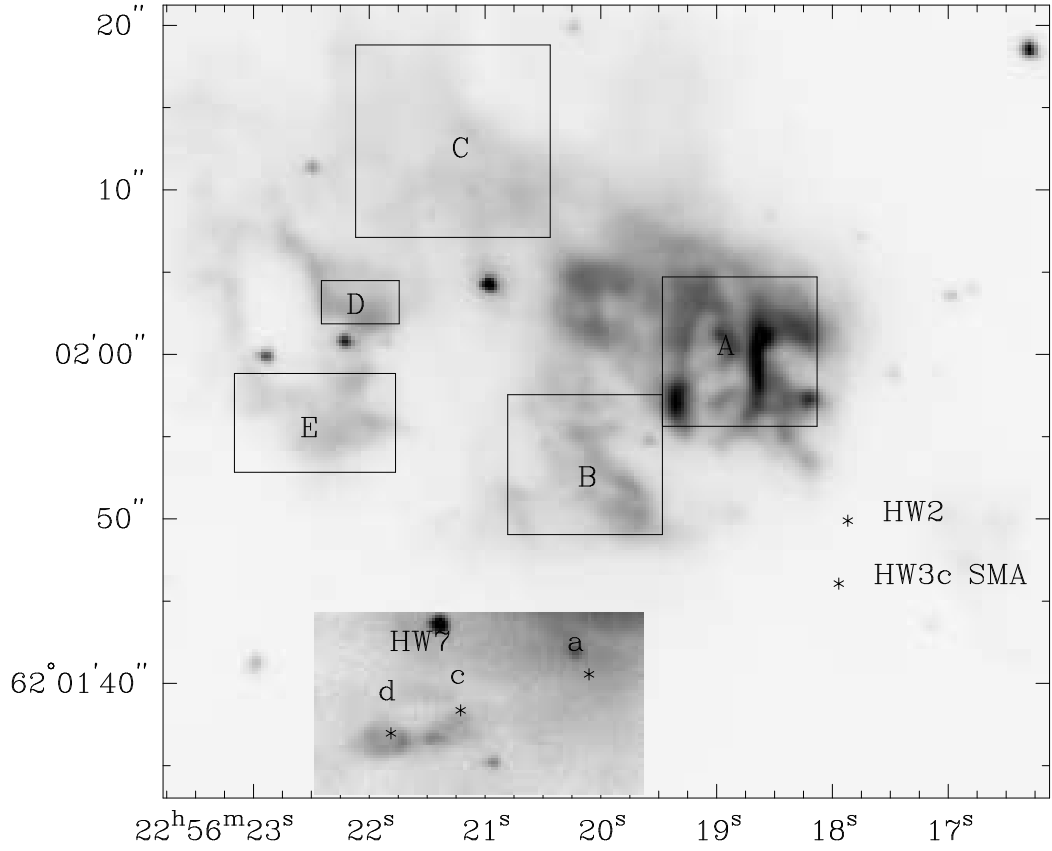}
\caption{
Image in the K' band of the central region of Cep~A East
taken in 2004 with linear intensity scaling, emphasizing the brightest
features.
The small rectangle containing the HW7 bipolar nebula is shown in a
different intensity scaling to properly show this much fainter object.
}
\end{figure}

\clearpage
\begin{figure}
\figurenum{2}
\includegraphics[scale=0.9,angle=0]{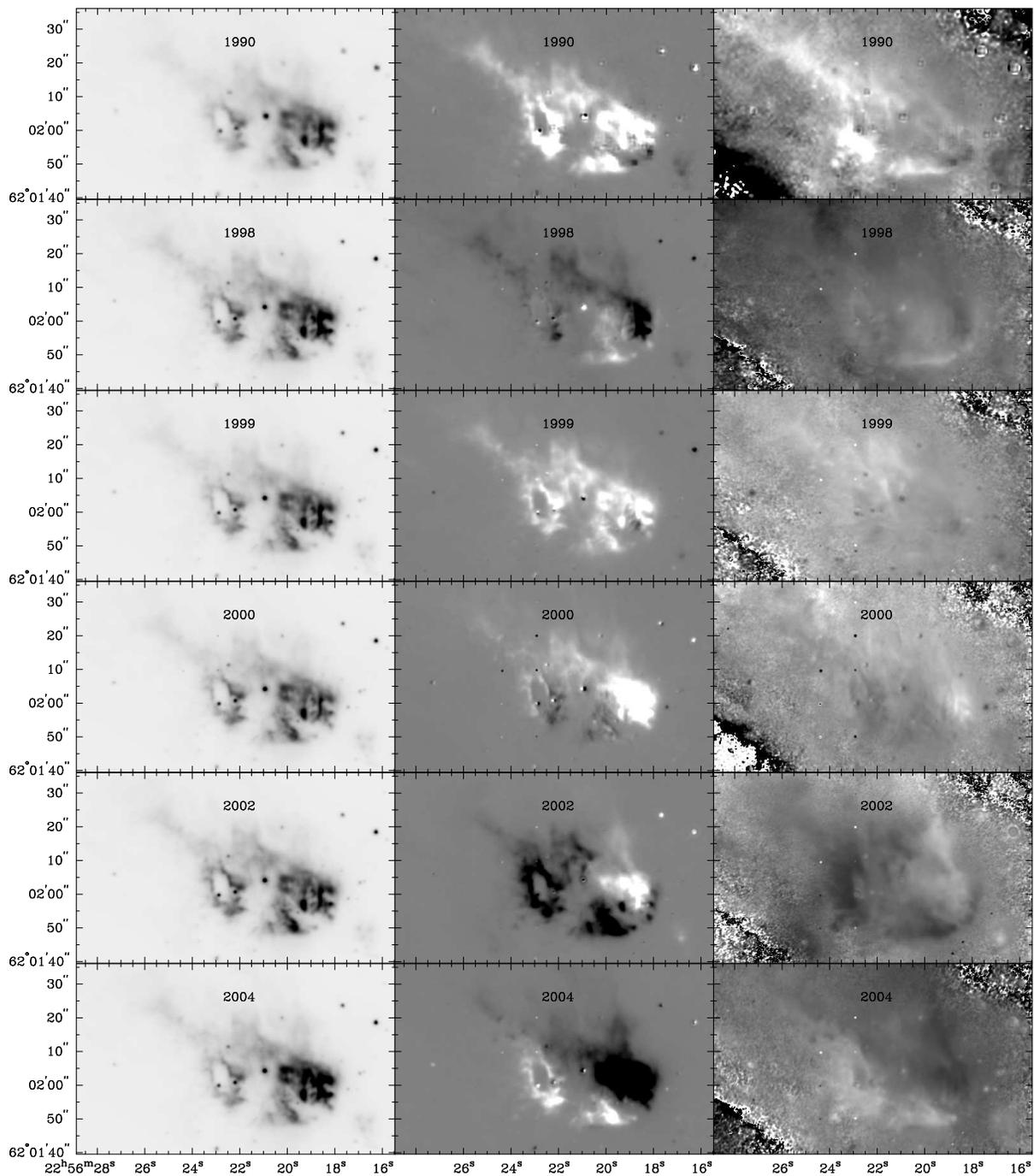}
\caption{
The left column shows the K' images of the central region of Cep A East
taken at the times listed in Table 1. The center column shows the difference
of those images and the average of the QUIRC images taken between 1998 and 2004.
These difference images clearly show the variations in relative brightness of
different parts of the Cep A East reflection nebula. 
The right column shows the difference images divided by the original image,
to separate the shadow from the structure of the reflection nebula.
}
\end{figure}

\clearpage
\begin{figure}
\figurenum{3}
\includegraphics[scale=0.8,angle=0]{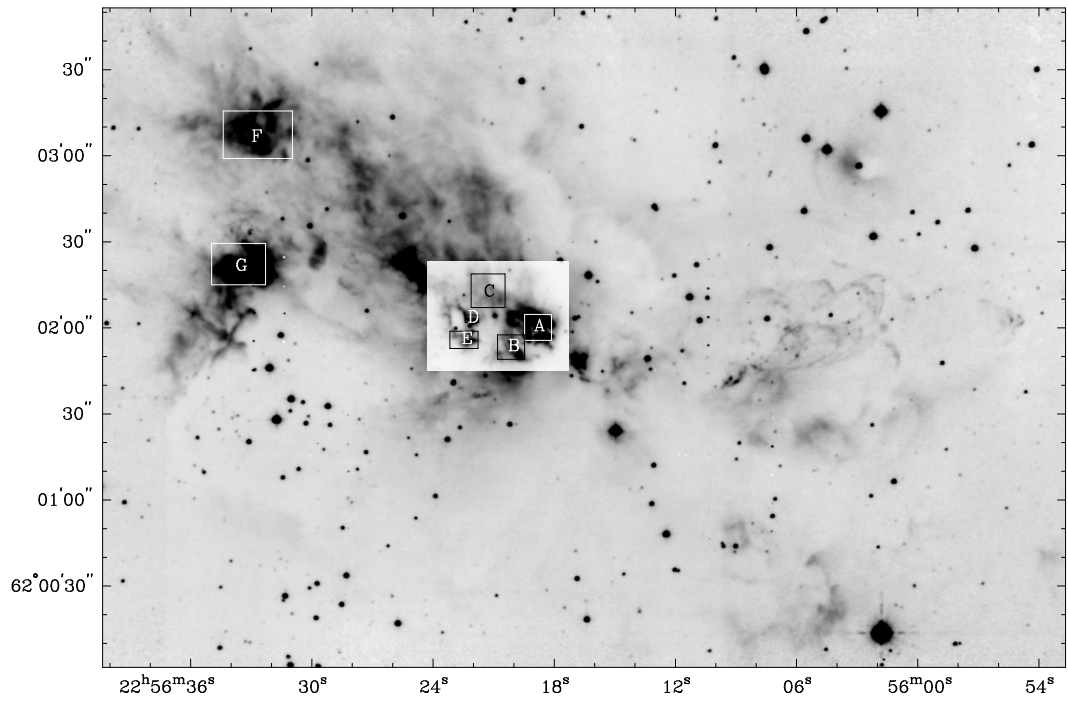}
\caption{
Image in the K' band of the Cep A East reflection nebula
taken in 1998
with integration boxes indicated and labeled.
The intensity scale is logarithmic to bring out the faint
features of the reflection nebula.
The light curves measured in those integration boxes are
shown in Fig.~3.
}
\end{figure}

\clearpage
\begin{figure}
\figurenum{4}
\includegraphics[scale=0.8,angle=0]{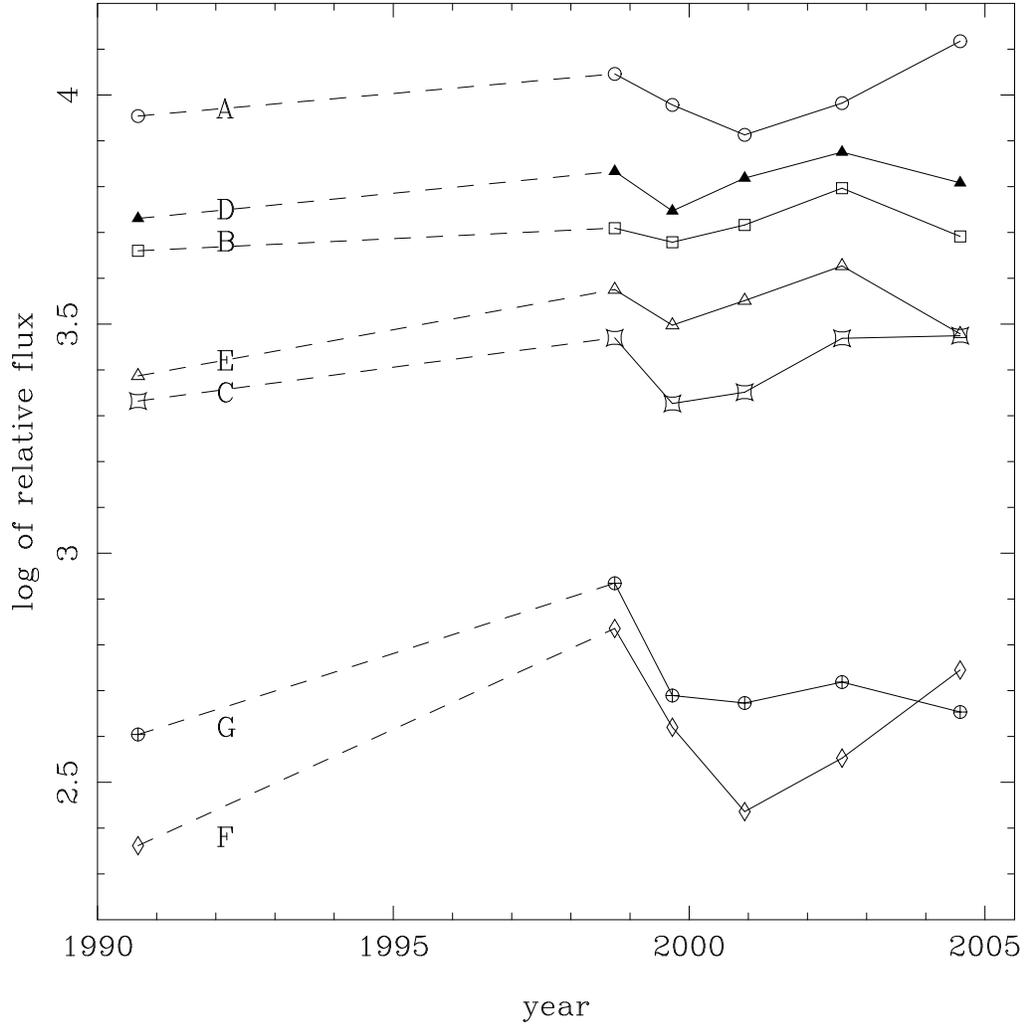}
\caption{
Light curves of the selected regions in the Cep A reflection nebula (see Fig.~2),
shown in logarithmic relative flux units.
}
\end{figure}

\clearpage
\begin{figure}
\figurenum{5}
\includegraphics[scale=0.8,angle=0]{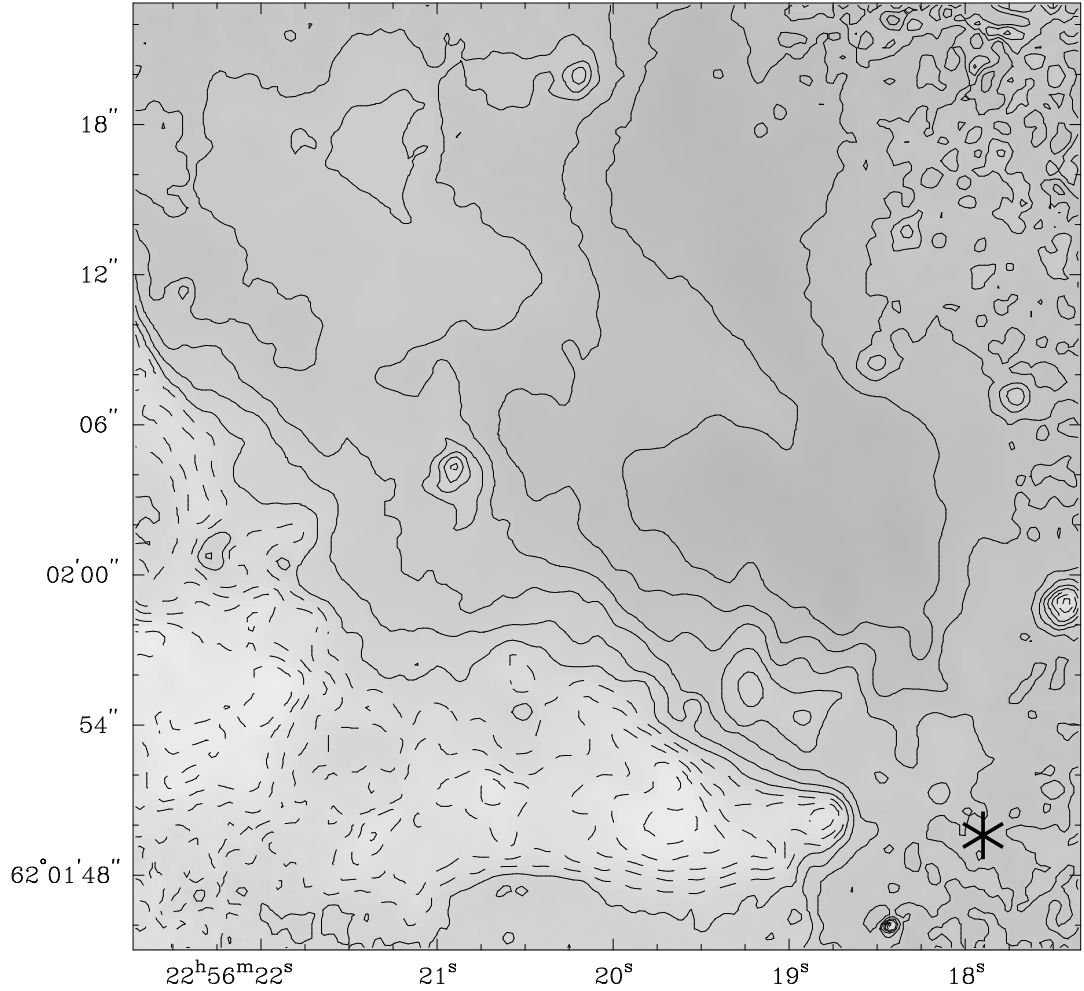}
\caption{
Contours of the normalized difference image
of the 2004 data, the same data as Fig. 2, lower right panel,
superposed on a grey-scale representation of the same data. 
Dashed contours indicate negative signals, solid lines
are positive signal contours.
This figure illustrates
that the shadow edge is a rather smooth transition.
It also shows that the shadow is sharper (steeper gradient)
close to the illuminating source HW2, which is indicated
by a star symbol.
}
\end{figure}

\clearpage
\begin{figure}
\figurenum{6}
\includegraphics[scale=0.8,angle=0]{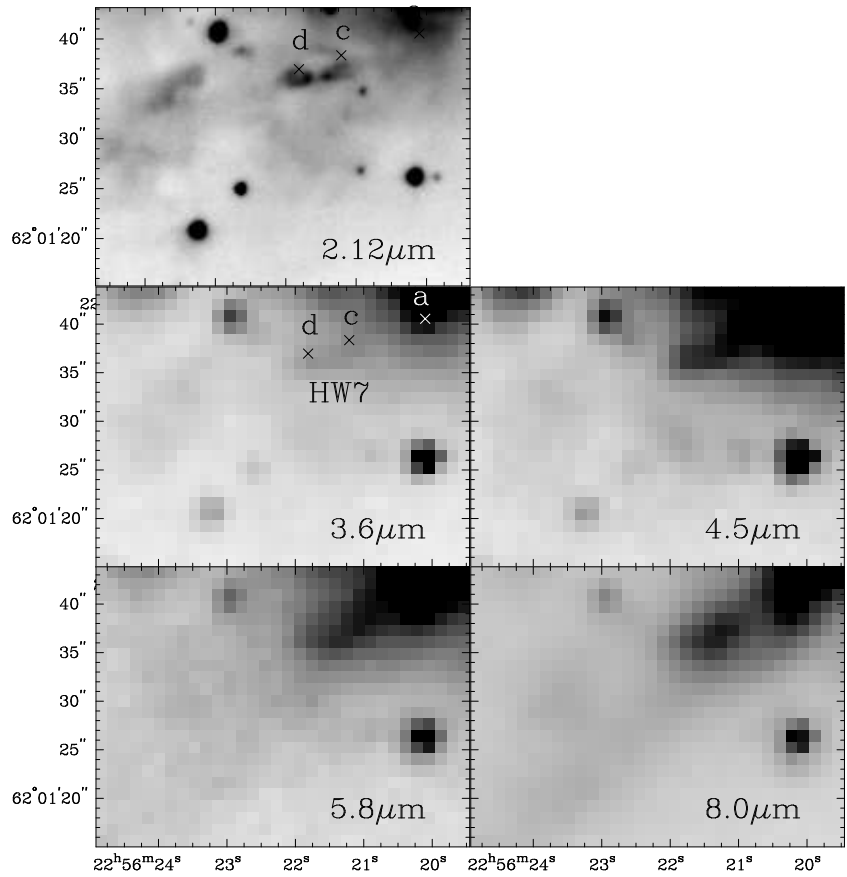}
\caption{
Comparison of our combined K' image and images
in the four Spitzer IRAC bands of the bipolar
nebula associated with HW7. The HW7 knots a, b, and c
are indicated by ''x'' symbols.
}
\end{figure}

\end{document}